\documentclass[preprint,showpacs,preprintnumbers,amsmath,amssymb]{revtex4}

\usepackage{graphicx}
\usepackage{dcolumn}
\usepackage{bm}

\begin{document}

\preprint{SCIENCE CHINA Physics,Mechanics \& Astronomy, Doi:10.1007/s11433-011-4320-2}

\title{Superallowed Fermi transitions in RPA with a relativistic point-coupling energy functional}

\author{Z. X. Li}
\author{J. M. Yao}%
 \email{Second.Author@institution.edu}
\author{H. Chen}
\affiliation{School of Physical Science and Technology, Southwest University, Chongqing 400715, China }

\date{\today}

\begin{abstract}
    The self-consistent random phase approximation (RPA) approach with the residual interaction derived from
    a relativistic point-coupling energy  functional is applied to evaluate the isospin symmetry-breaking corrections
    $\delta_c $ for the $0^+ \rightarrow 0^+$ superallowed Fermi transitions. With these $\delta_c$ values,
    together with the available experimental $ f t $ values and the improved radiative corrections, the
    unitarity of the Cabibbo-Kobayashi-Maskawa (CKM) matrix is
    examined. Even with the consideration of uncertainty, the sum of squared top-row elements has been shown to
    deviate from the unitarity condition by $~0.1\%$ for all the employed relativistic energy functionals.
\end{abstract}

\maketitle

\section{Introduction}

    In the standard model, the Cabibbo-Kobayashi-Maskawa (CKM) matrix~\cite{Cabibbo63, Kobayashi73} relates the quark eigenstates of
    the weak interaction with the quark mass eigenstates, and therefore it should be unitary. Examination of the unitarity of the CKM matrix
    provides a rigorous test for the standard model description of electroweak interactions. Usually, much attention
    has been paid to the top row of the CKM matrix. The unitarity requires, $\sum_{f=d,s,b}\vert V_{uf}\vert^2=1$,
    where the values of $\vert V_{us}\vert $ and $\vert V_{ub}\vert $ can be taken from the Particle Data Group reviews~\cite{Amsler08}.
    The value of matrix element $\vert V_{ud}\vert $ can be determined in nuclear physics through the
    following ways: nuclear $ 0^+ \rightarrow 0^+ $ superallowed Fermi transition~\cite{Hardy05},
    neutron decay \cite{Thompson90}, pion $\beta$ decay \cite{Pocanic04} and nuclear mirror transition~\cite{Naviliat-Cuncic09}.
    Among them, the first method provides the most precise determination of $ \vert V_{u d} \vert $ \cite{Amsler08},
    given by $\vert G_V/G_F\vert$, where $G_F$ is the Fermi coupling constant for purely leptonic decays.
    The vector coupling constant of semileptonic weak interactions $G_V$ is determined by the nucleus-independent $\mathcal Ft$
    value and transition-independent part of radiative corrections. Therefore, many efforts are devoted into the study of
    the nucleus-independent $\mathcal Ft$ value.

    On the theoretical side, there are several nuclear structure models that have been adopted to calculate the nucleus-independent
    $\mathcal Ft$ value, including the shell model~\cite{Towner08} and the self-consistent charge-exchange random phase approximation (RPA)
    based on both non-relativistic~\cite{Sagawa96} and relativistic~\cite{Liang08,Liang09} energy density functionals, as well as the
    isospin-projection scheme on top of the Skyrme-DFT approach~\cite{Satula10}. It has been shown
    in Ref.~\cite{Liang09} that the constancy of the $\mathcal Ft$ values can be satisfied for all the employed relativistic
    meson-exchange energy functionals.

    Recently, as the counterpart of meson-exchange energy functional, the relativistic point-coupling energy
    functional has attracted more and more attention because of its simplicity and success. In the point-coupling functional,
    there is no mesonic degree of freedom that makes its practical applications more feasible and the numerical effort
    considerably reduced~\cite{Zhao10}. In particular, the point-coupling functional can be easily extended to study
    the effects beyond the mean-field approximation that are important for a proper description of
    the nuclear low-lying collective excited states~\cite{Tamara06A, Tamara06B, Yao10}.

    In view of these facts, recently, the model of change-exchange RPA with the residual interaction
    derived from the relativistic density-dependent point-coupling functional has been developed~\cite{Finelli07}.
    In this work, we will adopt the similar model, but we'll start from the non-linear version of point-coupling functional to study
    the isospin symmetry-breaking corrections $\delta_c $ in the $0^+ \rightarrow 0^+$ superallowed transitions.
    With these $\delta_c$ values, together with
    the most recent experimental $ft$ values \cite{Hardy09} and the improved radiative corrections~\cite{Towner08, Marciano06}, the
    unitarity of the CKM matrix will be examined.

    This paper is arranged as follows. The model is introduced briefly in Sec.~\ref{sec2}. The results and corresponding discussions
    are given in Sec.~\ref{sec3}. A brief summary of the present investigation is presented in Sec.~\ref{sec4}.

 \section{\label{sec2}The model}

    For the charge-exchange channels of both the $\tau_+$ and $\tau_-$, the relativistic RPA equation has the following form~\cite{Liang09},
    \begin{eqnarray}
        \label{Equation: RPA}
        \begin{pmatrix}
             \mathcal A_{p \bar n, p' \bar n'}^J & \mathcal B_{p \bar n, n' \bar p'}^J \\
            -\mathcal B_{n \bar p, p' \bar n'}^J &-\mathcal A_{n \bar p, n' \bar p'}^J
        \end{pmatrix}
        \begin{pmatrix}
            U_{p' \bar n'}^{J \nu} \\
            V_{n' \bar p'}^{J \nu}
        \end{pmatrix}
        = \omega_\nu
        \begin{pmatrix}
            U_{p \bar n}^{J \nu} \\
            V_{n \bar p}^{J \nu}
        \end{pmatrix}.
    \end{eqnarray}
    In the above equation, $\bar p $ and $p $ ($\bar n $ and $n$) denote occupied and unoccupied proton (neutron)
    states, where the unoccupied states include the states above the Fermi surface and those in the Dirac sea.
    $U^{J \nu}$ and $V^{J \nu}$ are the amplitudes corresponding to the RPA energy $\omega_\nu$.
    The matrices $\mathcal A$ and $\mathcal B$ are given by,
    \begin{eqnarray}
        \label{Matrix A: RPA}
        \mathcal A_{12, 34}
            & =&   (E_1-E_2)\delta_{12, 34} + \langle 1 4 \vert V_{\rm ph} \vert 3 2\rangle, \\
        \label{Matrix B: RPA}
        \mathcal B_{12, 34}
            & =& - \langle 1 3 \vert V_{\rm ph} \vert 4 2 \rangle,
    \end{eqnarray}
    where $E_1, E_2$ are the single-particle energies from mean-field calculations.

    For the sake of the self-consistency, the particle-hole residual interaction $ V_{\rm ph} $ is derived from the same
    effective Lagrangian density as the mean-field Dirac single-nucleon Hamiltonian that determines the mean-field
    for the nuclear ground-state. The spin-isospin-dependent interaction terms are generated by the isovector terms.
    Although the direct one-$\pi$ contribution to the nuclear ground-state vanishes at the Hartree level because of
    parity conservation, the pion nevertheless must be included in the calculation of spin-isospin excitations~\cite{Paar04}.

    The derivative type of the pion-coupling necessitates the inclusion of the zero-range Landau-Migdal term, which
    accounts for the contact part of the nucleon-nucleon interaction,
    \begin{equation}
    \label{LM}
        V_{\delta\pi}(1, 2)
            = g^\prime [\dfrac{f_\pi}{m_\pi} \gamma_0\gamma_5 \mathbf{\gamma} \vec\tau]_1 [\dfrac{f_\pi}{m_\pi}
            \gamma_0\gamma_5\mathbf{\gamma} \vec\tau]_2 \delta(\mathbf{r}_1-\mathbf{r}_2),
    \end{equation}
    where $f^2_\pi/4\pi=0.08$, $m_\pi=138$ MeV. The parameter $g^\prime$ in principle has to be adjusted to reproduce the
    experimental Gamow-Teller resonance excitation energy~\cite{Paar04}. However, the direct contribution from the $\pi$-meson
    field vanishes for $0^+\rightarrow0^+$ transition. Therefore, the self-consistency is still maintained in the RH+RPA study
    of superallowed Fermi transition and we take $g^\prime=1/3$ as Ref.~\cite{Liang09}.

    In the calculations, the relativistic point-coupling energy functional of both PC-F1~\cite{Burvenich02} and
    PC-PK1~\cite{Zhao10} is adopted, where the scalar-isovector coupling terms do not exist. As a result, the particle-hole residual
    interaction $V_{\rm ph}$ that gives non-zero contribution to the charge-exchange RPA matrix is composed of three parts:
    the vector-isovector coupling terms~\cite{Niksic05},
    \begin{equation}
     V^{TV}(1, 2) =  [\gamma_0\gamma^\mu\vec\tau]_1(\alpha_{TV}+\delta_{TV}\Delta) [\gamma_0\gamma_\mu\vec\tau]_2 \delta(\mathbf{r}_1-\mathbf{r}_2),
     \end{equation}
    the Landau-Migdal term (\ref{LM}), and $\pi$-meson field
    term~\cite{Liang09,Paar04},
    \begin{equation}
     V_{\pi}(1, 2)= -[\dfrac{f_\pi}{m_\pi} \gamma_0\gamma_5 \mathbf{\gamma}\cdot\nabla \vec\tau ]_1
     [\dfrac{f_\pi}{m_\pi} \gamma_0\gamma_5  \mathbf{\gamma}\cdot\nabla\vec\tau ]_2
     D_\pi(\mathbf{r}_1,\mathbf{r}_2),
     \end{equation}
    where $D_\pi(\mathbf{r}_1,\mathbf{r}_2)$ denotes the $\pi$-meson propagator,
    \begin{equation}
    D_\pi(\mathbf{r}_1,\mathbf{r}_2) = \dfrac{1}{4\pi} \dfrac{e^{-m_\pi\vert \mathbf{r}_1-\mathbf{r}_2\vert}}{\vert
    \mathbf{r}_1-\mathbf{r}_2\vert}.
    \end{equation}
    $\alpha_{TV}$ and $\delta_{TV}$ are the coupling constants in the vector-isovector channel
    of point-coupling energy functional~\cite{Zhao10,Burvenich02}, and $\Delta$ is the Laplace operator.
    $\gamma^\mu$ are the four-component Dirac matrices. At the Hartree level, the Coulomb term does not contribute to the charge-exchange
    RPA matrix element.

    The eigenvectors of the RPA Eq.~(\ref{Equation: RPA}) are separated into two groups, which respectively represent the
    excitations of the $ \tau_- $ and $ \tau_+ $ channels with the following normalization conditions
    \begin{eqnarray}
        \sum\limits_{p \bar n} ( U_{p \bar n}^{J \nu} )^2 - \sum\limits_{n \bar p} ( V_{n \bar p}^{J \nu} )^2
            &=& + 1, ~ \text{ for $\tau_-$ channel }, \\
        \sum\limits_{p \bar n} ( U_{p \bar n}^{J \nu} )^2 - \sum\limits_{n \bar p} ( V_{n \bar p}^{J \nu} )^2
            &=& - 1, ~ \text{ for $\tau_+$ channel }.
    \end{eqnarray}
    The excitation energies $\Omega_\nu$ and the corresponding forward $X^{J \nu}$, backward $Y^{J \nu}$ amplitudes
    in the $\tau_- $ and $ \tau_+ $ channels are given by,
    \begin{eqnarray}
         \Omega_\nu = + \omega_\nu, ~X_{p \bar n}^{J \nu} = U_{p \bar n}^{J \nu},
        ~Y_{n \bar p}^{J \nu} = V_{n \bar p}^{J \nu}, ~ \text{ for $\tau_-$ channel }, \\
         \Omega_\nu = - \omega_\nu, ~X_{n \bar p}^{J \nu} = V_{n \bar p}^{J \nu},
        ~Y_{p \bar n}^{J \nu} = U_{p \bar n}^{J \nu}, ~ \text{ for $\tau_+$ channel }.
    \end{eqnarray}
    Subsequently, it is straightforward to calculate the $0^+ \rightarrow 0^+ $ superallowed transition probabilities
    between the ground-state and excited states with the forward and backward amplitudes for the $\tau_- $ and $ \tau_+ $
    channels as follows,
    \begin{eqnarray}
        \hspace{-5mm}
        B_{J \nu}^-
            & = & \left | \sum\limits_{p \bar n} X_{p \bar n}^{J \nu} \langle p \parallel \tau_- \parallel \bar n \rangle
                        + \sum\limits_{n \bar p} \kappa_{n\bar p} Y_{n \bar p}^{J \nu} \langle \bar p \parallel \tau_- \parallel n \rangle \right |^2, \\
        \hspace{-5mm}
        B_{J \nu}^+
            & = & \left | \sum\limits_{n \bar p} X_{n \bar p}^{J \nu} \langle n \parallel \tau_+ \parallel \bar p \rangle
                        + \sum\limits_{p \bar n} \kappa_{p\bar n} Y_{p \bar n}^{J \nu} \langle \bar n \parallel \tau_+ \parallel p \rangle \right |^2,
    \end{eqnarray}
    with $\kappa_{ab}=(-)^{j_a + j_b}$.

  \section{\label{sec3}Results and discussions}

    For simplicity, the pairing correlations are neglected and the filling approximation is used. Moreover, the calculation is presently
    restricted to preserve spherical symmetry. As a result, the Dirac equation for nucleons can be solved easily in coordinate space
    using the numerical techniques in Ref.~\cite{Meng98}, where the box size $R=15$ fm and the mesh size $\Delta r = 0.1$ fm.
    The solutions of Dirac equation, including single-particle wave functions and energies, are used as inputs of the RPA equation.
    As usual, the single-particle energy truncation is introduced. We find that with the choice of the truncations
    $[-940, 1100]$~MeV for the PC-F1 and $[-940, 1160]$~MeV for the PC-PK1, the model-independent sum rule
    \begin{eqnarray}
        \sum\limits_\nu B_\nu^- - \sum\limits_\nu B_\nu^+ = N - Z
    \end{eqnarray}
    can be fulfilled up to $ 10^{-5} $ accuracy, and the isospin symmetry-breaking corrections $ \delta_c $ are stable with respect to
    these numerical inputs at the same level of accuracy.

    \begin{figure}[ht]
        \centering
        \includegraphics[width=12cm]{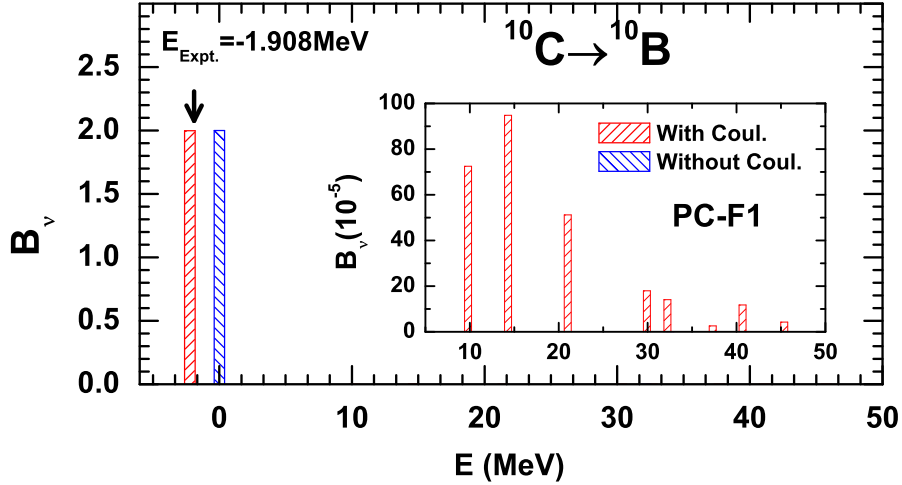}
        \caption{\label{fig1}Distribution of the strengths $B_\nu(0^+ \rightarrow 0^+_\nu)$ for the superallowed transition
                             $^{10}$C $\rightarrow$ $^{10}$B.}
    \end{figure}

    \begin{table}[ht]
        \centering
        \tabcolsep=12pt
        \caption{\label{tab1}Isospin symmetry-breaking corrections $ \delta_c $ (in \%) for the $0^+ \rightarrow 0^+$
                             superallowed transitions obtained by RPA with the relativistic point-coupling energy functional
                             of both PC-F1 and PC-PK1. The results are compared with those obtained with the relativistic
                             meson-exchange energy functionals PKO1 and DD-ME2~\cite{Liang09}.}
        \begin{tabular*}{110mm}{                      r @{\hspace*{1mm}}      c @{\hspace*{1mm}}      l
                                @{\extracolsep{\fill}}c @{\extracolsep{\fill}}c @{\extracolsep{\fill}}c @{\extracolsep{\fill}}c }
            \hline\hline
                                      &                 &                      & PC-F1 & PC-PK1 & PKO1~\cite{Liang09}
                                                                                                         & DD-ME2~\cite{Liang09} \\
            \hline
                $ ^{10}\textup  C   $ & $ \rightarrow $ & $^{10}\textup  B   $ & 0.109 & -      & 0.082  & 0.150  \\
                $ ^{14}\textup  O   $ & $ \rightarrow $ & $^{14}\textup  N   $ & 0.150 & -      & 0.114  & 0.197  \\
                $ ^{18}\textup {Ne} $ & $ \rightarrow $ & $^{18}\textup  F   $ & 0.309 & 0.297  & 0.270  & 0.430  \\
                $ ^{26}\textup {Si} $ & $ \rightarrow $ & $^{26}\textup {Al} $ & 0.202 & 0.180  & 0.176  & 0.252  \\
                $ ^{30}\textup  S   $ & $ \rightarrow $ & $^{30}\textup  P   $ & 0.420 & 0.488  & 0.497  & 0.633  \\
                $ ^{34}\textup {Ar} $ & $ \rightarrow $ & $^{34}\textup {Cl} $ & 0.379 & 0.378  & 0.268  & 0.376  \\
                $ ^{38}\textup {Ca} $ & $ \rightarrow $ & $^{38}\textup  K   $ & 0.347 & 0.325  & 0.313  & 0.441  \\
                $ ^{42}\textup {Ti} $ & $ \rightarrow $ & $^{42}\textup {Sc} $ & 0.400 & 0.375  & 0.384  & 0.523  \\
            \hline
                $ ^{26}\textup {Al} $ & $ \rightarrow $ & $^{26}\textup {Mg} $ & 0.159 & 0.141  & 0.139  & 0.198  \\
                $ ^{34}\textup {Cl} $ & $ \rightarrow $ & $^{34}\textup  S   $ & 0.316 & 0.314  & 0.234  & 0.307  \\
                $ ^{38}\textup  K   $ & $ \rightarrow $ & $^{38}\textup {Ar} $ & 0.294 & 0.275  & 0.278  & 0.371  \\
                $ ^{42}\textup {Sc} $ & $ \rightarrow $ & $^{42}\textup {Ca} $ & 0.345 & 0.322  & 0.333  & 0.448  \\
                $ ^{54}\textup {Co} $ & $ \rightarrow $ & $^{54}\textup {Fe} $ & 0.339 & 0.301  & 0.319  & 0.393  \\
                $ ^{66}\textup {As} $ & $ \rightarrow $ & $^{66}\textup {Ge} $ & 0.522 & 0.488  & 0.475  & 0.572  \\
                $ ^{70}\textup {Br} $ & $ \rightarrow $ & $^{70}\textup {Se} $ & 0.935 & 0.998  & 1.140  & 1.268  \\
                $ ^{74}\textup {Rb} $ & $ \rightarrow $ & $^{74}\textup {Kr} $ & 0.668 & 0.597  & 1.088  & 1.258  \\
            \hline\hline
        \end{tabular*}
    \end{table}

    The isospin symmetry-breaking correction $ \delta_c $ is determined by the superallowed transition strength $M^2_{F}$ as follows,
    \begin{equation}
        M^2_{F} \equiv \vert\langle f\vert \tau_\pm\vert i\rangle\vert^2
            = \vert M_0\vert^2(1-\delta_c),
    \end{equation}
    where $M_0=\sqrt{2}$ for $T=1$ states with the exact isospin symmetry, and $\vert f\rangle, \vert i\rangle$ represent the final
    and initial states of $\tau_\pm$ transitions respectively. It has to be pointed out that in practical calculations, there are
    several final states $0^+_\nu$ with different excitation energies $E_\nu$ and strengths $B_\nu$ as shown in Figure ~\ref{fig1} for
    the superallowed transition, for example, $^{10}$C $\rightarrow$ $^{10}$B. As usual, the final state $\vert f\rangle$ is taken to be the
    excited state with the largest transition strength $B_\nu$ (close to 2). Table~\ref{tab1} presents the isospin symmetry-breaking
    correction $\delta_c $ in several $ 0^+_i \rightarrow 0^+_f $ superallowed transitions, calculated using the RPA for the
    relativistic point-coupling energy functional of both PC-F1 and PC-PK1. For comparison, the results in Ref.\cite{Liang09}
    from the RPA calculations but with the relativistic meson-exchange energy functionals PKO1~\cite{Long06} and
    DD-ME2~\cite{Lalazissis05} are given as well. It has been found in Ref.~\cite{Liang09} that
    the significant differences in $\delta_c$ values obtained by PKO1 and by DD-ME2 are mainly due to the different treatments of
    the Coulomb field. The Fock terms have been included in the former case (PKO1), but not in the latter case (DD-ME2).
    Table~\ref{tab1} shows that the point-coupling functionals PC-F1 and PC-PK1 without the Fock terms
    give quite similar $\delta_c$ values, both of which are in between the results by the PKO1 and DD-ME2.

    \begin{table}[ht]
        \centering
        \tabcolsep=12pt
        \caption{\label{tab2}Excitation energies (in MeV) for the $ 0^+ \rightarrow 0^+ $ superallowed transitions
                             measured by taking the ground-state of the corresponding even-even nuclei as reference.
                             For comparison with the experimental values taken from the recent survey \cite{Hardy09},
                             the corrections due to the proton-neutron mass difference in particle-hole configurations
                             are made for the calculated results.}
        \begin{tabular*}{120mm}{                      r @{\hspace*{1mm}}      c @{\hspace*{1mm}}      l
                                @{\extracolsep{\fill}}c @{\extracolsep{\fill}}c @{\extracolsep{\fill}}c @{\extracolsep{\fill}}c
                                @{\extracolsep{\fill}}c }
            \hline\hline
                                      &                 &                      & Expt.  & PC-F1  & PC-PK1 & PKO1~\cite{Liang09}
                                                                                                                   & DD-ME2~\cite{Liang09} \\
            \hline
                $ ^{10}\textup  C   $ & $ \rightarrow $ & $^{10}\textup  B   $ & -1.908 & -2.217 & -      & -1.698 & -2.236 \\
                $ ^{14}\textup  O   $ & $ \rightarrow $ & $^{14}\textup  N   $ & -2.831 & -2.967 & -      & -2.420 & -3.081 \\
                $ ^{18}\textup {Ne} $ & $ \rightarrow $ & $^{18}\textup  F   $ & -3.402 & -3.400 & -3.419 & -3.195 & -3.451 \\
                $ ^{26}\textup {Si} $ & $ \rightarrow $ & $^{26}\textup {Al} $ & -4.842 & -4.956 & -5.055 & -4.531 & -5.110 \\
                $ ^{30}\textup  S   $ & $ \rightarrow $ & $^{30}\textup  P   $ & -5.460 & -5.295 & -5.330 & -4.845 & -5.395 \\
                $ ^{34}\textup {Ar} $ & $ \rightarrow $ & $^{34}\textup {Cl} $ & -6.063 & -5.975 & -5.964 & -5.559 & -6.278 \\
                $ ^{38}\textup {Ca} $ & $ \rightarrow $ & $^{38}\textup  K   $ & -6.612 & -6.582 & -6.576 & -6.035 & -6.775 \\
                $ ^{42}\textup {Ti} $ & $ \rightarrow $ & $^{42}\textup {Sc} $ & -7.000 & -6.833 & -6.869 & -6.661 & -6.964 \\
            \hline
                $ ^{26}\textup {Al} $ & $ \rightarrow $ & $^{26}\textup {Mg} $ &  4.233 &  4.211 &  4.304 &  3.908 &  4.350 \\
                $ ^{34}\textup {Cl} $ & $ \rightarrow $ & $^{34}\textup  S   $ &  5.492 &  5.292 &  5.288 &  5.062 &  5.561 \\
                $ ^{38}\textup  K   $ & $ \rightarrow $ & $^{38}\textup {Ar} $ &  6.044 &  5.905 &  5.905 &  5.557 &  6.083 \\
                $ ^{42}\textup {Sc} $ & $ \rightarrow $ & $^{42}\textup {Ca} $ &  6.426 &  6.207 &  6.242 &  6.118 &  6.333 \\
                $ ^{54}\textup {Co} $ & $ \rightarrow $ & $^{54}\textup {Fe} $ &  8.244 &  8.016 &  8.122 &  7.720 &  8.240 \\
                $ ^{66}\textup {As} $ & $ \rightarrow $ & $^{66}\textup {Ge} $ &  9.579 &  9.367 &  9.434 &  9.044 &  9.677 \\
                $ ^{70}\textup {Br} $ & $ \rightarrow $ & $^{70}\textup {Se} $ &  9.970 &  9.735 &  9.806 &  9.632 &  9.852 \\
                $ ^{74}\textup {Rb} $ & $ \rightarrow $ & $^{74}\textup {Kr} $ & 10.417 & 10.246 & 10.299 & 10.005 & 10.437 \\
            \hline\hline
        \end{tabular*}
    \end{table}

    In Table~\ref{tab2}, the excitation energies $E_\nu$ for the $ 0^+_i \rightarrow 0^+_f $ superallowed transitions are shown.
    For comparison with the experimental values taken from the recent survey~\cite{Hardy09}, the corrections due to
    the proton-neutron mass difference in particle-hole configurations are made for the calculated results. It is shown that
    all the relativistic energy functionals can reproduce the excitation energies quite well.

    The nucleus-independent $\mathcal Ft$ can be obtained by the experimental $ft$ value, isospin symmetry-breaking
    correction $\delta_c$ and transition-independent part of radioactive corrections~\cite{Hardy05},
    \begin{eqnarray}
        \mathcal F t
            = f t ( 1 + \delta^\prime_{R} ) ( 1 + \delta_{N S} - \delta_c ),
    \end{eqnarray}
    where $ \delta^\prime_{R}$ is the part of nucleus-independent radiative correction, which is a functional only of
    the electron's energy and the charge of daughter nucleus $Z$ while $\delta_{N S}$ is the part of radiative correction
    that depends on the details of nuclear structure.

    With the recent theoretical results of $\delta^\prime_{R} $ and $ \delta_{NS} $~\cite{Towner08} and the recent
    experimental $ f t $ values~\cite{Hardy09}, the nucleus-independent $\mathcal Ft_i$ value and its uncertainty
    $\delta \mathcal Ft_i$ for each superallowed Fermi transition are obtained with the calculated $ \delta_c $ values and
    listed in Table~\ref{tab3}, where the corresponding average value $\overline{\mathcal Ft}$ and $\chi^2/\nu$ are given as well.
    In the calculation of $\delta \mathcal Ft_i$, the uncertainty of $ \delta_c $ is taken as zero. The average value
    $\overline{\mathcal Ft}$ is calculated in the following way~\cite{Eilelman04},
    \begin{equation}
        \label{Average: Ft}
        \overline{\mathcal Ft} \pm \delta \overline{\mathcal Ft}
            = \frac{\sum_i \omega_i \mathcal Ft_i}{\sum_i \omega_i} \pm (\sum\limits_i \omega_i)^{-1/2}
    \end{equation}
    where the weight is
    \begin{equation}
        \omega_i
            = 1/(\delta \mathcal Ft_i)^2.
    \end{equation}
    The $\chi^2/\nu$ is calculated by
    \begin{equation}
        \chi^2/\nu
            = \sum^N\limits_i (\frac{\mathcal Ft_i - \overline{\mathcal Ft}}{\delta \mathcal Ft_i})^2/(N-1)
    \end{equation}
    where $N$ is the number of the calculated superallowed Fermi transitions. Since the uncertainties  $\delta \mathcal Ft_i$
    in the superallowed Fermi transitions  $^{34}$Ar $\rightarrow$ $^{34}$Cl and $^{74}$Rb $\rightarrow$ $^{74}$Kr~ are
    obviously larger than those in other transitions and excluded in the calculations of average value $\overline{\mathcal Ft}$
    and $\chi^2/\nu$.

    \begin{table}[ht]
        \centering
        \tabcolsep=12pt
        \caption{\label{tab3}Nucleus-independent $ \mathcal F t $ values (in s), its average value
                             $ \overline{\mathcal F t}$ (in s) and the $\chi^2 / \nu $. }
        \begin{tabular*}{80mm}{                      r @{\hspace*{1mm}}      c @{\hspace*{1mm}} l
                               @{\extracolsep{\fill}}c @{\extracolsep{\fill}}c }
            \hline\hline
                                      &                 &                       & PC-F1      & PC-PK1     \\
            \hline
                $ ^{10}\textup  C   $ & $ \rightarrow $ & $ ^{10}\textup  B   $ & 3078.7(45) & -          \\
                $ ^{14}\textup  O   $ & $ \rightarrow $ & $ ^{14}\textup  N   $ & 3077.0(31) & -          \\
                $ ^{34}\textup {Ar} $ & $ \rightarrow $ & $ ^{34}\textup {Cl} $ & 3078.5(84) & 3078.5(84) \\
                $ ^{26}\textup {Al} $ & $ \rightarrow $ & $ ^{26}\textup {Mg} $ & 3077.0(13) & 3077.6(13) \\
                $ ^{34}\textup {Cl} $ & $ \rightarrow $ & $ ^{34}\textup  S   $ & 3081.0(15) & 3081.1(15) \\
                $ ^{38}\textup  K   $ & $ \rightarrow $ & $ ^{38}\textup {Ar} $ & 3083.6(16) & 3084.2(16) \\
                $ ^{42}\textup {Sc} $ & $ \rightarrow $ & $ ^{42}\textup {Ca} $ & 3082.3(21) & 3083.0(21) \\
                $ ^{54}\textup {Co} $ & $ \rightarrow $ & $ ^{54}\textup {Fe} $ & 3083.3(24) & 3084.4(24) \\
                $ ^{74}\textup {Rb} $ & $ \rightarrow $ & $ ^{74}\textup {Kr} $ & 3119.6(88) & 3120.2(88) \\
                \hline
                 \multicolumn{3}{c}{average}                                 & 3080.3(7)  & 3081.1(7)     \\
                 \multicolumn{3}{c}{$\chi^2/\nu$}                            &    1.1     &    1.4        \\
            \hline\hline
        \end{tabular*}
    \end{table}

    Table~\ref{tab3} shows that the $\chi^2/\nu$ is 1.1 for PC-F1 and 1.4 for PC-PK1. It indicates that the constancy of
    the ${\mathcal Ft}$ values is good for both cases.

    \begin{figure}[ht]
        \centering
        \includegraphics[width=10cm]{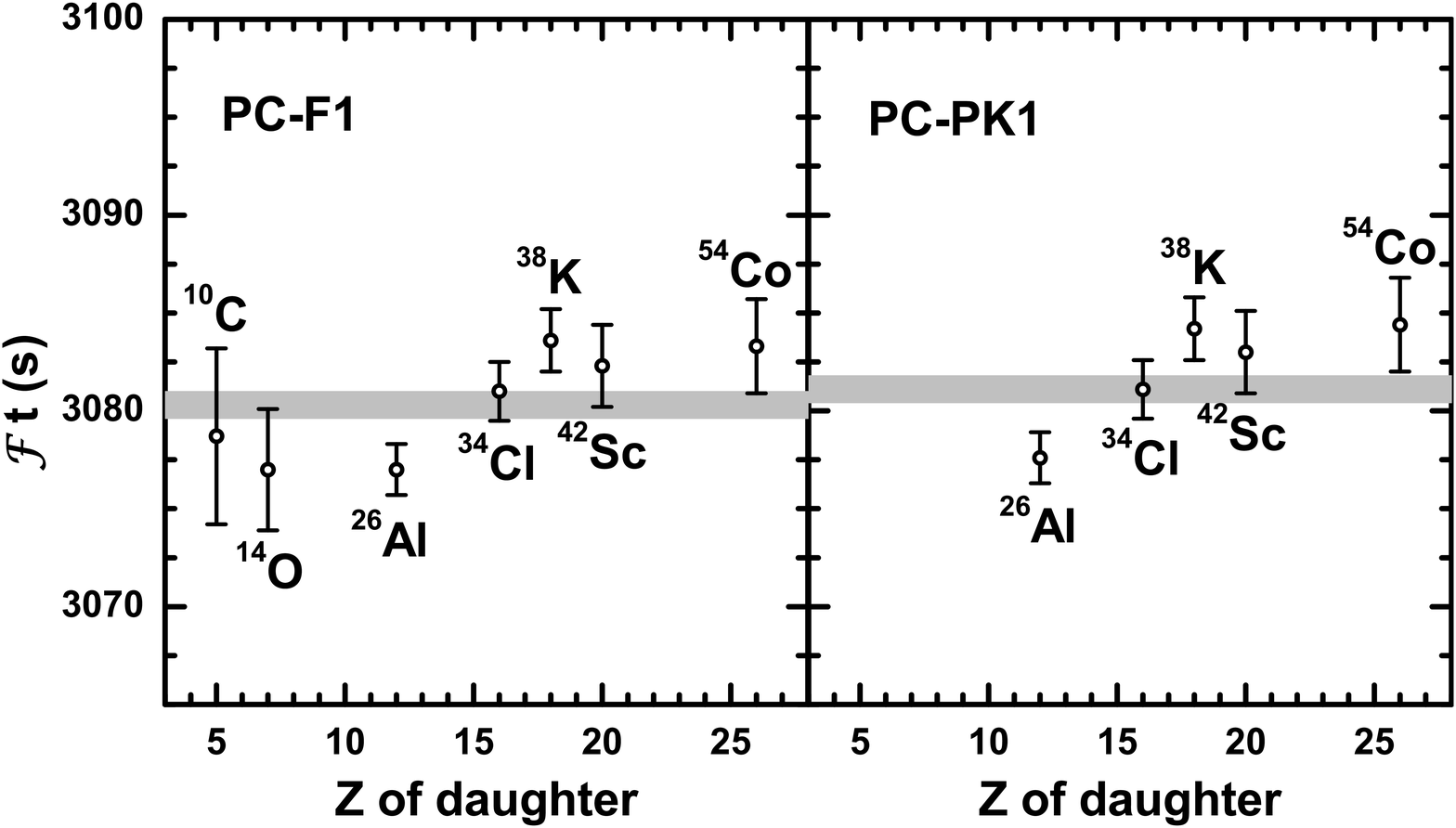}
        \caption{\label{fig2}Nucleus-independent $ \mathcal F t $ values as a function of the charge of
                             the daughter nucleus $Z$. The shaded horizontal band gives one standard deviation
                             around the average $\overline{\mathcal Ft}$ value.}
    \end{figure}

    To illustrate the constancy of the $\mathcal Ft$ values more clearly, we plot the nucleus-independent $\mathcal Ft$ values
    as a function of the charge of the daughter nucleus $Z$ in Figure~\ref{fig2}. The shaded horizontal band gives one standard
    deviation around the average $\overline{\mathcal Ft}$ value.

    With the nucleus-independent $\overline {\mathcal F t} $ value, the absolute value of element $V_{u d}$ in
    the CKM matrix can be calculated by
    \begin{eqnarray}
        \vert V_{u d}\vert^2 = \frac{ K }{ 2 G_F^2 ( 1 + \Delta_R^V ) \overline {\mathcal F t} }
    \end{eqnarray}
    where $K / ( \hbar c )^6 = 8120.2787(11) \times 10^{-10} ~ \textup{GeV}^{-4} ~ \textup{s}$, the purely leptonic decays
    Fermi coupling constant $G_F / (\hbar c)^3 = 1.16637(1) \times 10^{-5} ~ \textup{GeV}^{-2}$ \cite{Amsler08} and
    the radiative corrections' transition-independent part $\Delta_R^V = 2.361(38) \%$~\cite{Towner08}.

    Together with the other two CKM matrix elements $\vert V_{us} \vert = 0.2255(19)$ and $\vert V_{ub} \vert = 0.00393(36)$~\cite{Amsler08},
    the unitarity of the CKM matrix can be examined by the sum of squared top-row elements of the CKM matrix. The values
    of $\vert V_{u d}\vert$ and  $\vert V_{ud} \vert^2 + \vert V_{us} \vert^2 + \vert V_{ub} \vert^2$ are listed in Table~\ref{tab4},
    where we also give the values of uncertainty, which might be underestimated to some extent as the uncertainty of $\delta_c$ was
    assumed to be zero and the systematic errors were not taken into account.

    \begin{table}[ht]
        \centering
        \tabcolsep=12pt
        \caption{\label{tab4}The absolute value of matrix element $V_{u d}$ and the sum of squared top-row elements of
                             the CKM matrix from the RPA calculations with different relativistic energy functionals.}
        \footnotesize
        \begin{tabular}{lcc}
            \hline\hline
                                      & $ \vert V_{ud} \vert $ & $ \vert V_{ud} \vert^2 + \vert V_{us} \vert^2 + \vert V_{ub} \vert^2 $ \\
            \hline
                PC-F1                 & 0.97290(21)            & 0.9974(10)                                                             \\
                PC-PK1                & 0.97278(22)            & 0.9972(10)                                                             \\
                PKO1~\cite{Liang09}   & 0.97273(27)            & 0.9971(10)                                                             \\
                DD-ME2~\cite{Liang09} & 0.97311(26)            & 0.9978(10)                                                             \\
            \hline\hline
        \end{tabular}
    \end{table}

    Table~\ref{tab4} shows that the $\vert V_{u d} \vert$ values obtained by PC-F1 and PC-PK1 are quite similar and close to the results
    of PKO1. In addition, it is seen that even with the uncertainty, the sum of squared top-row elements deviates from the unitarity
    condition by $~0.1\%$ in the charge-exchange RPA calculations with these four relativistic effective interactions.

 \section{\label{sec4}Summary}

    In summary, the self-consistent RPA with the residual interaction derived from the relativistic point-coupling energy functional
    has been applied to calculate the isospin symmetry-breaking corrections $ \delta_c $ for several typical $ 0^+ \rightarrow 0^+ $
    superallowed transitions. Together with the experimental $ ft $ values in the most recent survey and the improved radiative
    corrections, the corresponding nucleus-independent $\mathcal Ft$ values and matrix element $ \vert V_{u d} \vert $ have been
    calculated. It has been found that the $\vert V_{u d} \vert$ values obtained by PC-F1 and PC-PK1 are quite similar and
    close to the results of PKO1. However, even with the uncertainty, the sum of squared top-row elements has been shown to
    deviate from the unitarity condition by $~0.1\%$ for all the employed relativistic energy functionals. It indicates that other effects,
    including deformation and pairing correlations, would play important roles. Therefore, it is very interesting to study
    the $ 0^+ \rightarrow 0^+ $ superallowed transitions in the framework of deformed QRPA with a proper pairing force, for instance
    a separable pairing force~\cite{Tian09}. Of course, in this case, particle number projection and angular momentum projection
    are required to give good nucleon number and angular momentum for RPA states. Work along this direction is in progress.

\begin{acknowledgments}
    One of the authors (Z. X. Li) would like to thank Z. M. Niu for providing the charge-exchange RPA code
    and thank H. Z. Liang, J. Meng,  P. Ring and D. Vretenar for helpful discussions.
    This work has been supported by the National Natural Science Foundation of China under Grant No.
    10947013, the Fundamental Research Funds for the Central Universities under Grant No. XDJK2010B007
    and the SWU Initial Research Foundation Grant to Doctor (SWU109011).
\end{acknowledgments}

\end{document}